\documentclass[12pt]{article}
\usepackage{amsmath,amssymb,amsthm}
\usepackage{cite}

\def\tsum{\textstyle\sum\limits}
\def\tprod{\textstyle\prod\limits}
\def\nt{\notin}
\def\sl{\mathrm{sl}}
\def\MB{\mathbf}
\def\wh{\widehat}
\def\ket#1{\left|\,#1\right>}

\def\slacs#1{\setlength{\arraycolsep}{#1}}

\slacs{.4ex}

\begin{document}
\title{New Formula for the Eigenvectors of the Gaudin Model
in the $\sl(3)$ Case.}
\author{\v C. Burd\'{\i}k\,$^{1)}$ and O. Navr\'atil\,$^{2)}$}
\maketitle

\begin{center}
$^{1)}$\,Department of Mathematics, Czech Technical University,

Faculty of Nuclear Sciences and Physical Engineering,

Trojanova 13, 120 00 Prague 2, Czech Republic

\medskip

$^{2)}$\,Department of Mathematics, Czech Technical University,

Faculty of Transportation Sciences,

Na Florenci 25, 110 00 Prague, Czech Republic
\end{center}

\begin{abstract}
We propose new formulas for eigenvectors of the Gaudin model in
the $\sl(3)$ case. The central point of the construction is the
explicit form of some operator $\MB{P}$, which is used for
derivation of eigenvalues given by the formula
$$
\left|\,\MB{w}_1,\MB{w}_2\right)=
{\tsum_{n=0}^\infty}\,\frac{\MB{P}^n}{n!}\ket{\MB{w}_1,\MB{w}_2,0},
$$
where $\MB{w}_1$, $\MB{w}_2$ fulfil the standard well-know Bethe
Ansatz equations.
\end{abstract}

\section{Introduction}

In 1973 M. Gaudin \cite{Gaudin1,Gaudin2,Gaudin3} proposed a new
class of integrable quantum models. These models were first
formulated for the algebra $g=\sl(2)$ and studied intensively by
many authors. Let the generators $e$, $f$ and $h$ form a standard
basis in $g=\sl(2)$. Let
$(\lambda)=(\lambda^{(1)},\ldots,\lambda^{(N)})$ be a set of
dominant integral weight of $\sl(2)$. Consider the tensor product
$V_{(\lambda)}=V_{\lambda^{(1)}}\otimes\ldots\otimes
V_{\lambda^{(N)}}$ and associate a complex number~$z_i$ with each
factor $V_{\lambda^{(i)}}$ of this tensor product.

For any element $x \in \sl(2)$ denote $x^{(i)}$ the operator
$1\otimes\ldots\otimes x \otimes\ldots1$, which acts as $x$ on the
$i^{\mathrm{th}}$ factor of $V_{(\lambda)}$. The Gaudin
hamiltonians are mutually commuting operators
\begin{equation}
\label{Ham0} H_i=\sum_{j\ne i}
\frac{\frac12\,h^{(i)}h^{(j)}+e^{(i)}f^{(j)}+e^{(j)}f^{(i)}}{z_i-z_j}\,.
\end{equation}
One of the main problems in Gaudin's model is to find eigenvectors
and eigenvalues of these operators.

The Bethe Ansatz method \cite{Baxter,Faddeev,Gaudin2,Sklyanin} is
maybe the most effective method for solving the Gaudin model in the
$g=\sl(2)$ case. It is trivially sees that the tensor of the highest
vectors of $V_{\lambda^{(i)}}$'s $\ket{0}=v_{\lambda^{(1)}}\otimes
v_{\lambda^{(2)}} \otimes\ldots\otimes v_{\lambda^{(N)}}$ is an
eigenvector in $V_{\lambda}$. One constructs other eigenvectors by
acting on this vector by the operators
$$
F(w_r)={\tsum_{i=1}^N}\frac{f^{(i)}}{w_r-z_i}
$$
depending on auxiliary parameters $w_1$, $w_2$, \ldots, $w_n$. The
vectors obtained in this way are called Bethe vectors. These vectors
are eigenvectors of the Gaudin model if
$$
{\tsum_{i=1}^N}\frac{\lambda^{(i)}}{w_r-z_i}- {\tsum_{s\neq
r}}\frac2{w_r-w_s}=0 \qquad\mathrm{for}\quad r=1,\ldots,n\,
$$
is fulfilled. These equations are called Bethe Ansatz equations. In
Section 2 we will briefly repeat the derivation of this well know
calculation.

In the nineties of the last century Feigin, Frenkel and Reshetikhin
\cite{Feigin} proved the formulas for eigenvectors of the general
semisimple Lie algebra $g$. The eigenvectors should be constructed
by applying the operators $f^{(i)}_j$, $j=1,\ldots,l$ connected with
simple roots to the vacuum $\ket{0}$. If now $F_1(w_1)$ and
$F_2(w_2)$ do not commute, we are not able to find Bethe equations
for $F_1(w_1)F_2(w_2)\ket{0}$. So it is needed to add some extra
terms. The right formula can be extracted from solutions of the KZ
equation \cite{SheVar} (and in fact can be obtained as
quai-classical asymptotics of such solutions \cite{ReshVar}.)
\begin{equation}
\label{F1}
\ket{w_1^{i_1},w_2^{i_2},\ldots,w_m^{i_m}}={\tsum_{p=\{I^1,\ldots,I^N\}}}
{\tprod_{j=1}^N} \frac{f^{(j)}_{i_1^j}f^{(j)}_{i_2^j}\ldots
f^{(j)}_{i_a^j}}
{(w_{{i_1^j}}-w_{{i_2^j}})(w_{{i_2^j}}-w_{{i_3^j}})\ldots(w_{{i_a^j}}-z_j)}
\ket{0}.
\end{equation}
Here the summation is taken over all ordered partitions $I^1 \cup
I^2\cup\ldots\cup I^N$ of the set $\{1,\ldots,m\}$, where
$I^j=\{i^j_1,i^j_2,\ldots,i^j_{a_j}\}$.


The main result of our paper is formulated in Section 3. For
construction of eigenvectors we use not only the operators
$F_j(w)$ which are connected with simple roots but the operators
which are connected with non-simple roots.

Explicitly, we will study the case $g=\sl(3)$. In this case, the Lie
algebra has two simple roots and the corresponding generators are
$f_1$ and $f_2$ will define $F_1(w_1)$ and $F_2(w_2)$. For the
non-simple root and the generator $f_3=[f_2,f_1]$ we will define
$F_3(w)$ in a similar way. It should be  mentioned that in the above
construction such operators were not used. Again, as in the $\sl(2)$
case, the tensor of the highest vectors of  $V_{\lambda_j}$'s is an
eigenvector in $V_{\lambda}$. We will denote it by $\ket{0}$ and by
$$
\ket{\MB{w}_1,\MB{w}_2,\MB{w}_3}= F_1(w_{1,1})F_1(w_{1,2})\ldots
F_1(w_{1,k})F_2(w_{2,1})\ldots F_2(w_{2,\ell})F_3(w_{3,1})\ldots
F_3(w_{3,m})\ket{0}.
$$
We will define by $\MB{P}$ the linear mapping
\begin{eqnarray*}
\MB{P}\ket{0,\MB{w}_2,\MB{w}_3}&=&\MB{P}\ket{\MB{w}_1,0,\MB{w}_3}=0\,,\\
\MB{P}\ket{\MB{w}_1,\MB{w}_2,\MB{w}_3}&=&
{\sum_{r,s}}\frac{\ket{\MB{w}_1-w_{1,r},\MB{w}_2-w_{2,s},\MB{w}_3+w_{1,r}}}
{w_{2,s}-w_{1,r}}\,.
\end{eqnarray*}
Our main theorem shows that the vectors
$$
\left|\,\MB{w}_1,\MB{w}_2\right)=
{\tsum_{n=0}^\infty}\,\frac{\MB{P}^n}{n!}\ket{\MB{w}_1,\MB{w}_2,0}.
$$
are eigenvectors of the Gaudin model if the Bethe Ansatz equations
are fulfilled.

\section{The Gaudin model for $\sl(2)$}

The $\sl(2)$ Gaudin model was studied by many authors
\cite{Gaudin1,Gaudin2,Gaudin3,HKR,KKM2,MPR,P,RSTS,S1,S3,ST} from
different points of view. In this section, we will concentrate on
the well-know Bethe Ansatz method for finding eigenvectors and
eigenvalues \cite{Baxter,Faddeev,Gaudin2,Sklyanin}.

Let the generators $e$, $f$ and $h$ form a standard basis in
$\sl(2)$ which fulfil the commutations relations
\begin{equation}
\label{sl2} \bigl[h,e\bigr]=2e\,,\qquad \bigl[h,f\bigr]=-2f\,,
\qquad\bigl[e,f\bigr]=h.
\end{equation}
For the second order Casimir operator we obtain
$$
C=ef+fe+\tfrac12\,h^2\,.
$$
We will define
\begin{equation}
\label{Uop-sl2} F(u)={\tsum_{i=1}^N}\frac{f^{(i)}}{u-z_i}\,,\qquad
E(u)={\tsum_{i=1}^N}\frac{e^{(i)}}{u-z_i}\,,\qquad
H(u)={\tsum_{i=1}^N}\frac{h^{(i)}}{u-z_i}\,.
\end{equation}

The central point in the Gaudin model is played by the operator
\begin{equation}
\label{T-sl2} T(u)= \tfrac12\,\Bigl(E(u)F(u)+F(u)E(u)+
\tfrac12\,H^2(u)\Bigr)\,.
\end{equation}
It is possible to rewrite operator (\ref{T-sl2}) in the form
$$
T(u)={\tsum_{i=1}^N}\frac{H_i}{u-z_i}+
\tfrac12\,{\tsum_{i=1}^N}\frac{C^{(i)}}{(u-z_i)^2}\,,
$$
where $H_i$ are given by (\ref{Ham0}) and $C^{(i)}$ is the Casimir
operator acting on the $i^{\mathrm{th}}$ factor of
$V_{(\lambda)}$.

It is easy to show that from the commutation relations (\ref{sl2})
and the definitions (\ref{Uop-sl2}) we obtain for $u\neq w$
\begin{eqnarray*}
\bigl[E(u),E(w)\bigr]&=&\bigl[F(u),F(w)\bigr]=\bigl[H(u),H(w)\bigr]=0\,,\\
\bigl[E(u),F(w)\bigr]&=&-\frac{H(u)-H(w)}{u-w}\,,\\
\bigl[H(u),E(w)\bigr]&=&-2\,\frac{E(u)-E(w)}{u-w}\,,\\
\bigl[H(u),F(w)\bigr]&=&2\,\frac{F(u)-F(w)}{u-w}\,.
\end{eqnarray*}

For construction of the Gaudin model we will use the highest
representations
$$
e^{(i)}v_{\lambda^{(i)}}=0\,,\qquad
h^{(i)}v_{\lambda^{(i)}}=\lambda^{(i)} v_{\lambda^{(i)}}\,.
$$
It is easy to see that the vector
$\ket{0}=v_{\lambda^{(1)}}\otimes v_{\lambda^{(2)}}
\otimes\ldots\otimes v_{\lambda^{(N)}}$ is eigenvectors of $T(u)$
and the relations
$$
E(u)\ket{0}=0\,,\quad H(u)\ket{0}=\lambda(u)\ket{0}=
{\tsum_{i=1}^N}\frac{\lambda^{(i)}}{u-z_i}\ket{0},\quad
T(u)\ket{0}=\tau(u)\ket{0},
$$
where $\tau(u)=\tfrac14\,\lambda^2(u)-\tfrac12\,\lambda'(u)$ are
hold.

We fix notation
$$
\begin{array}{l}
F({\MB{w}})=F(w_1)F(w_2)\ldots F(w_n)\,,\\[4pt]
F({\MB{w}-w_r})=F(w_1)\ldots F(w_{r-1})F(w_{r+1})\ldots
F(w_n)\,,\\[4pt]
F({\MB{w}+u})=F(u)F(w_1)\ldots F(w_n)
\end{array}
$$
and we can try to obtain, in accordance with the Bethe Ansatz
method, further eigenvalues  $\ket{\MB{w}}=F({\MB{w}})\ket{0}$.

Direct calculation gives
\begin{eqnarray*}
\bigl[T(u),F(\MB{w})\bigr]&=&
-{\tsum_{r=1}^n}\frac{F(\MB{w})}{u-w_r}\Bigl(H(u)-
{\tsum_{s\neq r}}\frac1{u-w_s}\Bigr)+\\
&&+ {\tsum_{r=1}^n}\frac{F(\MB{w}+u-w_r)}{u-w_r}\Bigl(H(w_r)-
{\tsum_{s\neq r}}\frac2{w_r-w_s}\Bigr).
\end{eqnarray*}
Applying this equation to the hight vector $\ket{0}$ we obtain
\begin{equation}
\label{Tw}
T(u)\ket{\MB{w}}=T_0(u)\ket{\MB{w}}+T_1(u)\ket{\MB{w}}\,,
\end{equation}
where
\begin{eqnarray*}
T_0(u)\ket{\MB{w}}&=&\tau(u)\ket{\MB{w}}-
{\tsum_{r=1}^n}\Bigl(\lambda(u)- {\tsum_{s\neq
r}}\dfrac1{u-w_s}\Bigr)\frac{\ket{\MB{w}}}{u-w_r}\,,\\
T_1(u)\ket{\MB{w}}&=& {\tsum_{r=1}^n}\Bigl(\lambda(w_r)-
{\tsum_{s\neq r}}\frac2{w_r-w_s}\Bigr)
\frac{\ket{\MB{w}+u-w_r}}{u-w_r}\,.
\end{eqnarray*}
It is evident that $\ket{\MB{w}}$ is the eigenvector $T(u)$ for all
$u$ iff
$$
T_1(u)\ket{\MB{w}}=0\quad\mathrm{and}\quad
T_0(u)\ket{\MB{w}}=\tau(u;\MB{w})\ket{\MB{w}}.
$$
The first equation is equivalent to the Bethe equations
\begin{equation}
\label{B-sl2} \lambda(w_r)- {\tsum_{s\neq r}}\frac2{w_r-w_s}=0
\quad\mbox{for all}~~r=1,\,\ldots,\,n
\end{equation}
and the second condition gives corresponding eigenvalue
\begin{equation}
\label{vh-sl2} \tau(u;\MB{w})=
\tau(u)-{\tsum_{r=1}^n}\frac{\lambda(u)}{u-w_r}+ {\tsum_{r\neq
s}}\frac2{(u-w_r)(w_r-w_s)}\,.
\end{equation}

\section{The Gaudin model for $\sl(3)$}

The $\sl(3)$ Gaudin model was studied by many authors
\cite{SK,MSTV,Re} from different points of view. We will concentrate
again on finding eigenvectors and eigenvalues. We use the method
analogous to that in section~2 for the case $\sl(2)$. In this
chapter, we formulate the main result of our paper.

We will start with a basis in $\mathrm{gl}(3)$, $e_{ij}$,
$i,j=1,2,3$, where
$$
[e_{ij},e_{kl}]={\delta}_{jk}e_{il}-{\delta}_{li}e_{kj}\,.
$$
The standard basis in $\sl(3)$ is then given by
$$
\begin{array}{c}
e_1=e_{12}\,,\quad e_2=e_{23}\,,\quad e_3=e_{13}\,,\quad
f_1=e_{21}\,,\quad f_2=e_{32}\,,\quad f_3=e_{31}\,,\\[4pt]
h_1=[e_1,f_1]=e_{11}-e_{22} \quad\mathrm{and}\quad
h_2=[e_2,f_2]=e_{22}-e_{33}\,.
\end{array}
$$
The second order Casimir operator is defined by
$$
C_2=e_1f_1+e_2f_2+e_3f_3+f_1e_1+f_2e_2+f_3e_3+
\tfrac23\,\bigl(h_1^2+h_1h_2+h_2^2\bigr)\,.
$$
So the Gaudin Hamiltonian is defined as
\begin{eqnarray*}
I(u)&=&\tfrac12\,\bigl(F_1(u)E_1(u)+F_2(u)E_2(u)+F_3(u)E_3(u)+\\
&&\hskip10mm+
E_1(u)F_1(u)+E_2(u)F_2(u)+E_3(u)F_3(u)\bigr)+\\
&&+ \tfrac13\,\bigl(H_1^2(u)+H_1(u)H_2(u)+H_2^2(u)\bigr)\,,
\end{eqnarray*}
where
$$
X(u)={\tsum_{i=1}^N}\frac{x^{(i)}}{u-z_i}\,.
$$
Direct calculation gives for $u\neq w$ the relation
$$
\bigl[X(u),Y(w)\bigr]=-\frac{Z(u)-Z(w)}{u-w}
\quad\mathrm{iff}\quad [x,y]=z\,.
$$
To fix the notation we write
$$
F_j({\MB{w}_j})= F_j(w_{1,1})F_j(w_{1,2})\ldots F_j(w_{1,k_j}),
\quad
F(\MB{w}_1,\MB{w}_2,\MB{w}_3)=F_1(\MB{w}_1)F_2(\MB{w}_2)F_3(\MB{w}_3)
$$
and
$$
\ket{\MB{w}_1,\MB{w}_2,\MB{w}_3}=F(\MB{w}_1,\MB{w}_2,\MB{w}_3)\ket{0}.
$$

The main idea is to use these vectors for construction of
eigenvectors of $I(u)$. Any long but straightforward calculations
give
$$
I(u)\ket{\MB{w}_1,\MB{w}_2,\MB{w}_3}=
\Bigl(I_1(u)+I_2(u)+I_3(u)+I_0(u)\Bigr)\ket{\MB{w}_1,\MB{w}_2,\MB{w}_3},
$$
where
$$
\begin{array}{rclrcl}
I_1(u)&=&I_1^0(u)+I_1^-(u)\,,\qquad&
I_2(u)&=&I_2^0(u)+I_2^-(u)+I_2^+(u)\,,\\[4pt]
I_3(u)&=&I_3^0(u)+I_3^+(u)\,,\qquad& I_0(u)&=&I_0^0(u)+I_0^+(u)
\end{array}
$$
and
\begin{eqnarray*}
I_1^0(u)\ket{\MB{w}_1,\MB{w}_2,\MB{w}_3}&=&
{\tsum_r}\Bigl(\lambda_1(w_{1,r})-{\tsum_{\wh{r}\neq r}}
\dfrac2{w_{1,r}-w_{1,\wh{r}}}+
{\tsum_s}\dfrac1{w_{1,r}-w_{2,s}}-{\tsum_t}\dfrac1{w_{1,r}-w_{3,t}}\Bigr)\times\\
&&\hskip30mm\times
\dfrac{\ket{\MB{w}_1+u-w_{1,r},\MB{w}_2,\MB{w}_3}}{u-w_{1,r}}-\\
&&-
{\tsum_{r,s}}\dfrac{\ket{\MB{w}_1+u-w_{1,r},\MB{w}_2+w_{1,r}-w_{2,s},\MB{w}_3}}
{(u-w_{1,r})(w_{1,r}-w_{2,s})}-\\
&&-
{\tsum_{r,t}}\dfrac{\ket{\MB{w}_1+u-w_{1,r},\MB{w}_2,\MB{w}_3+w_{1,r}-w_{3,t}}}
{(u-w_{3,t})(w_{3,t}-w_{1,r})}\\
I_1^-(u)\ket{\MB{w}_1,\MB{w}_2,\MB{w}_3}&=&
-{\tsum_t}\dfrac{\ket{\MB{w}_1+u,\MB{w}_2+w_{3,t},\MB{w}_3-w_{3,t}}}{u-w_{3,t}}\\[12pt]
\end{eqnarray*}

\vfill\eject

\begin{eqnarray*}
I_2^0(u)\ket{\MB{w}_1,\MB{w}_2,\MB{w}_3}&=&
{\tsum_s}\Bigl(\lambda_2(w_{2,s})+
{\tsum_r}\dfrac1{u-w_{1,r}}-{\tsum_{\wh{s}\neq s}}
\dfrac2{w_{2,s}-w_{2,\wh{s}}}-
{\tsum_t}\dfrac1{u-w_{3,t}}\Bigr)\times\\
&&\hskip30mm\times
\dfrac{\ket{\MB{w}_1,\MB{w}_2+u-w_{2,s},\MB{w}_3}}{u-w_{2,s}}\\
I_2^-(u)\ket{\MB{w}_1,\MB{w}_2,\MB{w}_3}&=&
{\tsum_t}\dfrac{\ket{\MB{w}_1+w_{3,t},\MB{w}_2+u,\MB{w}_3-w_{3,t}}}{u-w_{3,t}}\\
I_2^+(u)\ket{\MB{w}_1,\MB{w}_2,\MB{w}_3}&=& -{\tsum_{r,s\neq\wh{s}}}
\dfrac{\ket{\MB{w}_1-w_{1,r},\MB{w}_2+u-w_{2,s}-w_{2,\wh{s}},\MB{w}_3+w_{1,r}}}
{(u-w_{1,r})(u-w_{2,s})(u-w_{2,\wh{s}})}\\[12pt]
I_3^0(u)\ket{\MB{w}_1,\MB{w}_2,\MB{w}_3}&=&
{\tsum_t}\Bigl(\lambda_3(w_{3,t})-
{\tsum_r}\dfrac1{w_{3,t}-w_{1,r}}-
{\tsum_s}\dfrac1{w_{3,t}-w_{2,s}}- {\tsum_{\wh{t}\neq t}}
\dfrac2{w_{3,t}-w_{3,\wh{t}}}\Bigr)\times\\
&&\hskip30mm\times
\dfrac{\ket{\MB{w}_1,\MB{w}_2,\MB{w}_3+u-w_{3,t}}}{u-w_{3,t}}-\\
&&-
{\tsum_{r,t}}\dfrac{\ket{\MB{w}_1+w_{3,t}-w_{1,r},\MB{w}_2,\MB{w}_3+u-w_{3,t}}}
{(u-w_{1,r})(w_{1,r}-w_{3,t})}+\\
&&+
{\tsum_{s,t}}\dfrac{\ket{\MB{w}_1,\MB{w}_2+w_{3,t}-w_{2,s},\MB{w}_3+u-w_{3,t}}}
{(u-w_{3,t})(w_{3,t}-w_{2,s})}\\
I_3^+(u)\ket{\MB{w}_1,\MB{w}_2,\MB{w}_3}&=&
{\tsum_{r,s}}\Bigl(\lambda_2(w_{1,r})-\lambda_2(w_{2,s})\Bigr)
\dfrac{\ket{\MB{w}_1-w_{1,r},\MB{w}_2-w_{2,s},\MB{w}_3+u}}
{(u-w_{1,r})(w_{1,r}-w_{2,s})}-\\
&&- 2{\tsum_{r,s\neq\wh{s}}}
\dfrac{\ket{\MB{w}_1-w_{1,r},\MB{w}_2-w_{2,s},\MB{w}_3+u}}
{(u-w_{1,r})(w_{1,r}-w_{2,\wh{s}})(w_{2,\wh{s}}-w_{2,s})}+\\
&&+ {\tsum_{r,s\neq\wh{s}}}
\dfrac{\ket{\MB{w}_1-w_{1,r},\MB{w}_2+w_{1,r}-w_{2,s}-w_{2,\wh{s}},\MB{w}_3+u}}
{(u-w_{1,r})(w_{1,r}-w_{2,s})(w_{1,r}-w_{2,\wh{s}})}\\[12pt]
I_0^0(u)\ket{\MB{w}_1,\MB{w}_2,\MB{w}_3}&=&
\tau(u)\ket{\MB{w}_1,\MB{w}_2,\MB{w}_3}-\\
&&- {\tsum_r}\Bigl(\lambda_1(u)-{\tsum_{\wh{r}\neq
r}}\dfrac2{w_{1,r}-w_{1,\wh{r}}}+ {\tsum_s}\dfrac1{w_{1,r}-w_{2,s}}-
{\tsum_t}\dfrac1{w_{1,r}-w_{3,t}}\Bigr)\times\\
&&\hskip30mm\times
\dfrac{\ket{\MB{w}_1,\MB{w}_2,\MB{w}_3}}{u-w_{1,r}}-\\
&&- {\tsum_s}\Bigl(\lambda_2(u)+
{\tsum_r}\dfrac1{w_{2,s}-w_{1,r}}-{\tsum_{\wh{s}\neq
s}}\dfrac2{w_{2,s}-w_{2,\wh{s}}}-
{\tsum_t}\dfrac1{w_{2,s}-w_{3,t}}\Bigr)\times\\
&&\hskip30mm\times
\dfrac{\ket{\MB{w}_1,\MB{w}_2,\MB{w}_3}}{u-w_{2,s}}-\\
&&- {\tsum_t}\Bigl(\lambda_3(u)- {\tsum_r}\dfrac1{w_{3,t}-w_{1,r}}-
{\tsum_s}\dfrac1{w_{3,t}-w_{2,s}}- {\tsum_{\wh{t}\neq t}}
\dfrac2{w_{3,t}-w_{3,\wh{t}}}\Bigr)\times\\
&&\hskip30mm\times
\dfrac{\ket{\MB{w}_1,\MB{w}_2,\MB{w}_3}}{u-w_{3,t}}+\\
&&+
{\tsum_{r,t}}\dfrac{\ket{\MB{w}_1+w_{3,t}-w_{1,r},\MB{w}_2,\MB{w}_3+w_{1,r}-w_{3,t}}}
{(u-w_{1,r})(u-w_{3,t})}\\
\end{eqnarray*}

\vfill\eject

\begin{eqnarray*}
I_0^+(u)\ket{\MB{w}_1,\MB{w}_2,\MB{w}_3}&=&
-{\tsum_{r,s}}\Bigl(\lambda_2(u)-\lambda_2(w_{2,s})\Bigr)\,
\dfrac{\ket{\MB{w}_1-w_{1,r},\MB{w}_2-w_{2,s},\MB{w}_3+w_{1,r}}}
{(u-w_{1,r})(u-w_{2,s})}+\\
&&+ 2{\tsum_{r,s\neq\wh{s}}}
\dfrac{\ket{\MB{w}_1-w_{1,r},\MB{w}_2-w_{2,s},\MB{w}_3+w_{1,r}}}
{(u-w_{1,r})(u-w_{2,\wh{s}})(w_{2,\wh{s}}-w_{2,s})}\,.
\end{eqnarray*}
In these formulas we use for simplicity
$\lambda_3(u)=\lambda_1(u)+\lambda_2(u)$.

Now we are able to construct the eigenvectors and calculate the
eigenvalues. We define a linear operator $\MB{P}$ by
\begin{eqnarray*}
\MB{P}\ket{0,\MB{w}_2,\MB{w}_3}&=&
\MB{P}\ket{\MB{w}_1,0,\MB{w}_3}=0\,,\\
\MB{P}\ket{\MB{w}_1,\MB{w}_2,\MB{w}_3}&=&
{\tsum_{r,s}}\frac{\ket{\MB{w}_1-w_{1,r},\MB{w}_2-w_{2,s},\MB{w}_3+w_{1,r}}}
{w_{2,s}-w_{1,r}}
\end{eqnarray*}
and  a vector
$$
\left|\,\MB{w}_1,\MB{w}_2\right)=
{\tsum_{n=0}^\infty}\,\frac{\MB{P}^n}{n!}\ket{\MB{w}_1,\MB{w}_2,0}
$$
for any $\MB{w}_1$ and $\MB{w}_2$. We show by the fulfilment of
Bethe Ansatz equations for $\lambda_1(w_{1,r})$ and
$\lambda_2(w_{2,s})$ that $\left|\,\MB{w}_1,\MB{w}_2\right)$ is the
eigenvector of $I(u)$ for any $u$. So we can formulate our main
theorem.

\bigskip

\noindent \textbf{Theorem.}

\smallskip

\noindent \textsl{If the Bethe Ansatz conditions: for any $r$
\begin{equation}
\label{B1} \lambda_1(w_{1,r})- {\tsum_{\wh{r}\neq r}}
\dfrac2{w_{1,r}-w_{1,\wh{r}}}+ {\tsum_s}\dfrac1{w_{1,r}-w_{2,s}}=0
\end{equation}
and for any $s$
\begin{equation}
\label{B2} \lambda_2(w_{2,s})+{\tsum_r}\dfrac1{w_{2,s}-w_{1,r}}-
{\tsum_{\wh{s}\neq s}} \dfrac2{w_{2,s}-w_{2,\wh{s}}}=0
\end{equation}
are fulfilled then the vector
$$
\left|\,\MB{w}_1,\MB{w}_2\right)=
{\tsum_{n=0}^\infty}\,\frac{\MB{P}^n}{n!}\ket{\MB{w}_1,\MB{w}_2,0}
$$
is the eigenvector of the $\sl(3)$ Gaudin model and
$$
I(u)\left|\,\MB{w}_1,\MB{w}_2\right)=
\tau(u;\MB{w}_1,\MB{w}_2)\left|\,\MB{w}_1,\MB{w}_2\right),
$$
where}
\begin{equation}
\label{VH3} \begin{array}{rcl}
\tau(u;\MB{w}_1,\MB{w}_2)&=&\tau(u)-
{\tsum_r}\Bigl(\lambda_1(u)-{\tsum_{\wh{r}\neq r}}
\dfrac2{w_{1,r}-w_{1,\wh{r}}}+
{\tsum_s}\dfrac1{w_{1,r}-w_{2,s}}\Bigr)\dfrac1{u-w_{1,r}}-\\[6pt]
&& \hskip10mm- {\tsum_s}\Bigl(\lambda_2(u)+{\tsum_r}
\dfrac1{w_{2,s}-w_{1,r}}- {\tsum_{\wh{s}\neq s}}
\dfrac2{w_{2,s}-w_{2,\wh{s}}}\Bigr)\dfrac1{u-w_{2,s}}\,.
\end{array}
\end{equation}

\noindent \textsc{Proof:} For $\left|\,\MB{w}_1,\MB{w}_2\right)$ to
be the eigenvector of $I(u)$ with eigenvalue
$\tau(u;\MB{w}_1,\MB{w}_2)$, there should be
\begin{equation}
\label{I1} I_1(u)\left|\,\MB{w}_1,\MB{w}_2\right)=
I_2(u)\left|\,\MB{w}_1,\MB{w}_2\right)=
I_3(u)\left|\,\MB{w}_1,\MB{w}_2\right)=0
\end{equation}
and
\begin{equation}
\label{I0} I_0(u)\left|\,\MB{w}_1,\MB{w}_2\right)=
\tau(u;\MB{w}_1,\MB{w}_2)\left|\,\MB{w}_1,\MB{w}_2\right).
\end{equation}
In the following Lemma~1 we show
$I_1(u)\left|\,\MB{w}_1,\MB{w}_2\right)=0$ in  detail. Since the
proof of the relations
$I_2(u)\left|\,\MB{w}_1,\MB{w}_2\right)=I_3(u)\left|\,\MB{w}_1,\MB{w}_2\right)=0$
is very similar, we will skip it. The proof of
$I_0(u)\left|\,\MB{w}_1,\MB{w}_2\right)=
\tau(u;\MB{w}_1,\MB{w}_2)\left|\,\MB{w}_1,\MB{w}_2\right)$ will be
given in Lemma~2.

\bigskip

\noindent \textbf{Lemma 1.}
\smallskip

\noindent \textsl{If  (\ref{B1}) is valid, then}
$I_1(u)\left|\,\MB{w}_1,\MB{w}_2\right)=0$.

\medskip

\noindent \textsc{Proof:}  It is easy to see that
$$
I_1^-(u)\ket{\MB{w}_1,\MB{w}_2,0}=0\,.
$$
For any $n\geq0$ we would like to have
$$
\Bigl(\tfrac1{n+1}\,I_1^-(u)\MB{P}^{n+1}+I_1^0(u)\MB{P}^n\Bigr)
\ket{\MB{w}_1,\MB{w}_2,0}=0\,
$$
valid.

For $n=0$ we have
$$
\begin{array}{l}
I_1^0(u)\ket{\MB{w}_1,\MB{w}_2,0}=\\[4pt]
\hskip10mm= {\tsum_{r_1}}\Bigl(\lambda_1(w_{1,r_1})- {\tsum_{r_2\neq
r_1}} \dfrac2{w_{1,r_1}-w_{1,r_2}}+
{\tsum_{s_1}}\dfrac1{w_{1,r_1}-w_{2,s_1}}\Bigr)
\dfrac{\ket{\MB{w}_1+u-w_{1,r_1},\MB{w}_2,0}}{u-w_{1,r_1}}-\\[6pt]
\hskip10mm- {\tsum_{r_1,s_1}}
\dfrac{\ket{\MB{w}_1+u-w_{1,r_1},\MB{w}_2+w_{1,r_1}-w_{2,s_1},0}}
{(u-w_{1,r_1})(w_{1,r_1}-w_{2,s_1})}\\[12pt]
I_1^-(u)\MB{P}\ket{\MB{w}_1,\MB{w}_2,0}=
-{\tsum_{r_1,s_1}}\dfrac{\ket{\MB{w}_1+u-w_{1,r_1},\MB{w}_2+w_{1,r_1}-w_{2,s_1},0}}
{(u-w_{1,r_1})(w_{2,s_1}-w_{1,r_1})}\,.
\end{array}
$$
For
$$
\Bigl(I_1^-(u)\MB{P}+I_1^0(u)\Bigr)\ket{\MB{w}_1,\MB{w}_2,0}=0\,
$$
to be valid we obtain condition (\ref{B1}) for any $r$
$$
\lambda_1(w_{1,r})-
{\tsum_{\wh{r}\neq r}} \dfrac2{w_{1,r}-w_{1,\wh{r}}}+
{\tsum_s}\dfrac1{w_{1,r}-w_{2,s}}=0\,.
$$
Here is the origin of the first Bethe Ansatz equation. The second
Bethe Ansatz equation arises in the same manner from the equation
$I_2(u)\left|\,\MB{w}_1,\MB{w}_2\right)=0$.

In order to simplify the notation, we denote by $R_n$ the ordered
set of numbers $r_1$, \ldots, $r_n$ where $r_i\neq r_k$ and
similarly $S_n$. So we can write
$$
\MB{P}^n\ket{\MB{w}_1,\MB{w}_2,0}=
{\tsum_{R_n,S_n}}\frac{\ket{\MB{w}_1-w_{1,r_1}-\ldots-w_{1,r_n},
\MB{w}_2-w_{2,s_1}-\ldots-w_{2,s_n},w_{1,r_1}+\ldots+w_{1,r_n}}}
{(w_{2,s_1}-w_{1,r_1})\ldots(w_{2,s_n}-w_{1,r_n})}\,,
$$
where the summation is over all sets $R_n$ and $S_n$. We will
abbreviate it as
$$
\MB{P}^n\ket{\MB{w}_1,\MB{w}_2,0}=
{\tsum_{R_n,S_n}}\frac{\ket{\MB{w}_1-w_{1,R_n},\MB{w}_2-w_{2,S_n},w_{1,R_n}}}{W_n}\,,\quad
\mathrm{where}\quad W_n={\tprod_{k=1}^n}(w_{2,s_k}-w_{1,r_k})\,.
$$
So we have
$$
\begin{array}{l}
I_1^0(u)\MB{P}^n\ket{\MB{w}_1,\MB{w}_2,0}=\\[4pt]
={\tsum_{R_n,S_n}}~{\tsum_{r_{n+1}\nt R_n}}
\Bigl(\lambda_1(w_{1,r_{n+1}})- {\tsum_{{r_{n+2}\nt
R_n}\atop{r_{n+2}\neq r_{n+1}}}}
\dfrac2{w_{1,r_{n+1}}-w_{1,r_{n+2}}}+ {\tsum_{s_{n+1}\nt
S_n}}\dfrac1{w_{1,r_{n+1}}-w_{2,s_{n+1}}}-\\[12pt]
\hskip20mm- {\tsum_{k=1}^n} \dfrac1{w_{1,r_{n+1}}-w_{1,r_k}}\Bigr)
\dfrac{\ket{\MB{w}_1+u-w_{1,R_n}-w_{1,r_{n+1}},\MB{w}_2-w_{2,S_n},w_{1,R_n}}}
{(u-w_{1,r_{n+1}})W_n}-\\[12pt]
\hskip10mm- {\tsum_{R_n,S_n}}~{\tsum_{{r_{n+1}\nt R_n}\atop
{s_{n+1}\nt S_n}}} \dfrac{\ket{\MB{w}_1+u-w_{1,R_n}-w_{1,r_{n+1}},
\MB{w}_2+w_{1,r_{n+1}}-w_{2,S_n}-w_{2,s_{n+1}},w_{1,R_n}}}
{(u-w_{1,r_{n+1}})(w_{1,r_{n+1}}-w_{2,s_{n+1}})W_n}-\\[12pt]
\hskip10mm- {\tsum_{R_n,S_n}}~{\tsum_{r_{n+1}\nt
R_n}}~{\tsum_{k=1}^n}
\dfrac{\ket{\MB{w}_1+u-w_{1,R_n}-w_{1,r_{n+1}},\MB{w}_2-w_{2,S_n},
w_{1,R_n}-w_{1,r_k}+w_{1,r_{n+1}}}}
{(u-w_{1,r_k})(w_{1,r_k}-w_{1,r_{n+1}})W_n}
\end{array}
$$
If we substitute $\lambda_1(w_{1,r_{n+1}})$ from (\ref{B1}) into the
above formula , we can adjust it to the form
$$
\begin{array}{l}
I_1^0(u)\MB{P}^n\ket{\MB{w}_1,\MB{w}_2,0}=\\[4pt]
= {\tsum_{R_n,S_n}}~{\tsum_{r_{n+1}\nt R_n}}~{\tsum_{k=1}^n}
\Bigl(\dfrac1{w_{1,r_{n+1}}-w_{1,r_k}}-\dfrac1{w_{1,r_{n+1}}-w_{2,s_k}}\Bigr)
\times\\[6pt]
\hskip20mm\times
\dfrac{\ket{\MB{w}_1+u-w_{1,R_n}-w_{1,r_{n+1}},\MB{w}_2-w_{2,S_n},w_{1,R_n}}}
{(u-w_{1,r_{n+1}})W_n}-\\[12pt]
\hskip5mm- {\tsum_{R_n,S_n}}~{\tsum_{{r_{n+1}\nt R_n}\atop
{s_{n+1}\nt S_n}}} \dfrac{\ket{\MB{w}_1+u-w_{1,R_n}-w_{1,r_{n+1}},
\MB{w}_2+w_{1,r_{n+1}}-w_{2,S_n}-w_{2,s_{n+1}},w_{1,R_n}}}
{(u-w_{1,r_{n+1}})(w_{1,r_{n+1}}-w_{2,s_{n+1}})W_n}-\\[12pt]
\hskip5mm- {\tsum_{R_n,S_n}}~{\tsum_{r_{n+1}\nt
R_n}}~{\tsum_{k=1}^n}
\dfrac{\ket{\MB{w}_1+u-w_{1,R_n}-w_{1,r_{n+1}},\MB{w}_2-w_{2,S_n},
w_{1,R_n}-w_{1,r_k}+w_{1,r_{n+1}}}}
{(u-w_{1,r_k})(w_{1,r_k}-w_{1,r_{n+1}})W_n}=\\[12pt]
={\tsum_{R_{n+1},S_n}}~{\tsum_{k=1}^n} \dfrac{(w_{2,s_k}-w_{2,r_k})
\ket{\MB{w}_1+u-w_{1,R_{n+1}},\MB{w}_2-w_{2,S_n},w_{1,R_{n+1}}-w_{1,r_{n+1}}}}
{(u-w_{1,r_{n+1}})(w_{1,r_{n+1}}-w_{1,r_k})(w_{2,s_k}-w_{1,r_{n+1}})W_n}-\\[12pt]
\hskip5mm- {\tsum_{R_{n+1},S_{n+1}}}
\dfrac{\ket{\MB{w}_1+u-w_{1,R_n}-w_{1,r_{n+1}},
\MB{w}_2+w_{1,r_{n+1}}-w_{2,S_n}-w_{2,s_{n+1}},w_{1,R_n}}}
{(u-w_{1,r_{n+1}})(w_{1,r_{n+1}}-w_{2,s_{n+1}})W_n}-\\[12pt]
\hskip5mm- {\tsum_{R_{n+1},S_n}}~{\tsum_{k=1}^n}
\dfrac{\ket{\MB{w}_1+u-w_{1,R_{n+1}},\MB{w}_2-w_{2,S_n},
w_{1,R_{n+1}}-w_{1,r_k}}}
{(u-w_{1,r_k})(w_{1,r_k}-w_{1,r_{n+1}})W_n}
\end{array}
$$
Since in the third element we sum over all possible combinations
$n+1$ of the different elements $r_i$, we can change in the sum
$r_k$ by $r_{n+1}$ and write the third element as
$$
-{\tsum_{R_{n+1},S_n}}~{\tsum_{k=1}^n}
\dfrac{(w_{2,s_k}-w_{1,r_k})\ket{\MB{w}_1+u-w_{1,R_{n+1}},\MB{w}_2-w_{2,S_n},
w_{1,R_{n+1}}-w_{1,r_n+1}}}
{(u-w_{1,r_{n+1}})(w_{1,r_{n+1}}-w_{1,r_k})(w_{2,s_k}-w_{1,r_{n+1}})W_n}\,,
$$
which cancels the first element. So we obtain
$$
I_1^0(u)\MB{P}^n\ket{\MB{w}_1,\MB{w}_2,0}= {\tsum_{R_{n+1},S_{n+1}}}
\dfrac{\ket{\MB{w}_1+u-w_{1,R_{n+1}},\MB{w}_2+w_{1,r_{n+1}}-w_{2,S_{n+1}},w_{1,R_n}}}
{(u-w_{1,r_{n+1}})W_{n+1}}\,.
$$
On the other hand, we have
$$
\begin{array}{l}
I_1^-(u)\MB{P}^{n+1}\ket{\MB{w}_1,\MB{w}_2,0}=\\[4pt]
\hskip20mm= -{\tsum_{R_{n+1},S_{n+1}}}~{\tsum_{k=1}^{n+1}}
\dfrac{\ket{\MB{w}_1+u-w_{1,R_{n+1}},\MB{w}_2+w_{1,r_k}-w_{2,S_{n+1}},
w_{1,R_{n+1}}-w_{1,r_k}}} {(u-w_{1,r_k})W_{n+1}}\,.
\end{array}
$$
Again we sum over all possible combinations $n+1$ of the different
elements $r_i$, we can change in the sum $r_k$ by $r_{n+1}$ and we
obtain
$$
\begin{array}{l}
I_1^-(u)\MB{P}^{n+1}\ket{\MB{w}_1,\MB{w}_2,0}=\\[4pt]
\hskip10mm= -{\tsum_{R_{n+1},S_{n+1}}}~{\tsum_{k=1}^{n+1}}
\dfrac{\ket{\MB{w}_1+u-w_{1,R_{n+1}},\MB{w}_2+w_{1,r_{n+1}}-w_{2,S_{n+1}},
w_{1,R_{n+1}}-w_{1,r_{n+1}}}}
{(u-w_{1,r_{n+1}})W_{n+1}}=\\[6pt]
\hskip10mm= -(n+1){\tsum_{R_{n+1},S_{n+1}}}
\dfrac{\ket{\MB{w}_1+u-w_{1,R_{n+1}},\MB{w}_2+w_{1,r_{n+1}}-w_{2,S_{n+1}},w_{1,R_n}}}
{(u-w_{1,r_{n+1}})W_{n+1}}=\\[12pt]
\hskip10mm= -(n+1)I_1^0(u)\MB{P}^n\ket{\MB{w}_1,\MB{w}_2,0}\,.
\end{array}
$$

\bigskip

\noindent \textbf{Lemma 2.}

\smallskip
\noindent \textsl{If  (\ref{B1}) and (\ref{B2}) is valid, then}
$I_0(u)\left|\,\MB{w}_1,\MB{w}_2\right)=
\tau(u;\MB{w}_1,\MB{w}_2)\left|\,\MB{w}_1,\MB{w}_2\right)$.

\medskip

\noindent \textsc{Proof:} It is easy to see that
$$
I_0^0(u)\ket{\MB{w}_1,\MB{w}_2,0}=
\tau(u;\MB{w}_1,\MB{w}_2)\ket{\MB{w}_1,\MB{w}_2,0}\,,
$$
where $\tau(u;\MB{w}_1,\MB{w}_2)$ is given in (\ref{VH3}).

Now we would like to prove for $n\geq1$ that
$$
\Bigl(\tfrac1{n!}\,\bigl(I_0^0(u)-\tau(u;\MB{w}_1,\MB{w}_2)\bigr)\MB{P}^n+
\tfrac1{(n-1)!}\,I_0^+(u)\MB{P}^{n-1}\Bigr)\ket{\MB{w}_1,\MB{w}_2,0}=0\,.
$$
If we start with the above expressions we obtain
$$
\begin{array}{l}
I_0^+(u)\MB{P}^{n-1}\ket{\MB{w}_1,\MB{w}_2,0}=
-{\tsum_{R_n,S_n}}\Bigl(\lambda_2(u)-\lambda_2(w_{2,s_n})\Bigr)
\dfrac{\ket{\MB{w}_1-w_{1,R_n},\MB{w}_2-w_{2,S_n},w_{1,R_n}}}
{(u-w_{1,r_n})(u-w_{2,s_n})W_{n-1}}+\\[9pt]
\hskip30mm+ 2{\tsum_{R_n,S_{n+1}}}
\dfrac{\ket{\MB{w}_1-w_{1,R_n},\MB{w}_2-w_{2,S_n},w_{1,R_n}}}
{(u-w_{1,r_n})(u-w_{2,s_{n+1}})(w_{2,s_{n+1}}-w_{2,n})W_{n-1}}\\[24pt]
I_0^0(u)\MB{P}^n\ket{\MB{w}_1,\MB{w}_2,0}= \tau(u){\tsum_{R_n,S_n}}
\dfrac{\ket{\MB{w}_1-w_{1,R_n},\MB{w}_2-w_{2,S_n},w_{1,R_n}}}{W_n}-\\[9pt]
\hskip15mm- {\tsum_{R_{n+1},S_n}}\Bigl(\lambda_1(u)-
{\tsum_{r_{n+2}\nt R_{n+1}}}\dfrac2{w_{1,r_{n+1}}-w_{1,r_{n+2}}}+
{\tsum_{s_{n+1}\nt S_n}}\dfrac1{w_{1,r_{n+1}}-w_{2,s_{n+1}}}-\\[6pt]
\hskip40mm- {\tsum_{k=1}^n}\dfrac1{w_{1,r_{n+1}}-w_{1,r_k}}\Bigr)
\dfrac{\ket{\MB{w}_1-w_{1,R_n},\MB{w}_2-w_{2,S_n},w_{1,R_n}}}
{(u-w_{1,r_{n+1}})W_n}-\\[9pt]
\hskip15mm- {\tsum_{R_n,S_{n+1}}}\Bigl(\lambda_2(u)+
{\tsum_{r_{n+1}\nt R_n}}\dfrac1{w_{2,s_{n+1}}-w_{1,r_{n+1}}}-
{\tsum_{s_{n+2}\nt S_{n+1}}}\dfrac2{w_{2,s_{n+1}}-w_{2,s_{n+2}}}-\\[6pt]
\hskip40mm- {\tsum_{k=1}^n}\dfrac1{w_{2,s_{n+1}}-w_{1,r_k}}\Bigr)
\dfrac{\ket{\MB{w}_1-w_{1,R_n},\MB{w}_2-w_{2,S_n},w_{1,R_n}}}
{(u-w_{2,s_{n+1}})W_n}-\\[9pt]
\end{array}
$$

$$
\begin{array}{l}
\hskip15mm- {\tsum_{R_n,S_n}}~{\tsum_{k=1}^n} \Bigl(\lambda_3(u)-
{\tsum_{r_{n+1}\nt R_n}}\dfrac1{w_{1,r_k}-w_{1,r_{n+1}}}-
{\tsum_{s_{n+1}\nt S_n}}\dfrac1{w_{1,r_k}-w_{2,s_{n+1}}}-\\[6pt]
\hskip40mm- {\tsum_{{\ell=1}\atop{\ell\neq k}}^n}
\dfrac2{w_{1,r_k}-w_{1,r_\ell}}\Bigr)
\dfrac{\ket{\MB{w}_1-w_{1,R_n},\MB{w}_2-w_{2,S_n},w_{1,R_n}}}
{(u-w_{1,r_k})W_n}+\\[9pt]
\hskip15mm+ {\tsum_{R_{n+1},S_n}}~{\tsum_{k=1}^n}
\dfrac{\ket{\MB{w}_1-w_{1,R_{n+1}}+w_{1,r_k},
\MB{w}_2-w_{2,S_n},w_{1,R_{n+1}}-w_{1,r_k}}}
{(u-w_{1,r_{n+1}})(u-w_{1,r_k})W_n}
\end{array}
$$
First we arrange the second, third and fourth elements in the
expression for $I_0^0(u)\MB{P}^n\ket{\MB{w}_1,\MB{w}_2,0}$ and then
change $r_k$ by $r_{n+1}$. So we obtain
$$
\begin{array}{l}
I_0^0(u)\MB{P}^n\ket{\MB{w}_1,\MB{w}_2,0}= \tau(u){\tsum_{R_n,S_n}}
\dfrac{\ket{\MB{w}_1-w_{1,R_n},\MB{w}_2-w_{2,S_n},w_{1,R_n}}}{W_n}-\\[6pt]
\hskip10mm- {\tsum_{R_{n+1},S_n}}\Bigl(\lambda_1(u)-
{\tsum_{r_{n+2}\neq r_{n+1}}}\dfrac2{w_{1,r_{n+1}}-w_{1,r_{n+2}}}+
{\tsum_{s_{n+1}}}\dfrac1{w_{1,r_{n+1}}-w_{2,s_{n+1}}}\Bigr)\times\\[6pt]
\hskip40mm\times
\dfrac{\ket{\MB{w}_1-w_{1,R_n},\MB{w}_2-w_{2,S_n},w_{1,R_n}}}
{(u-w_{1,r_{n+1}})W_n}-\\[6pt]
\hskip10mm- {\tsum_{R_{n+1},S_n}}~{\tsum_{k=1}^n}
\Bigl(\dfrac1{w_{1,r_{n+1}}-w_{1,r_k}}-
\dfrac1{w_{1,r_{n+1}}-w_{2,s_k}}\Bigr)
\dfrac{\ket{\MB{w}_1-w_{1,R_n},\MB{w}_2-w_{2,S_n},w_{1,R_n}}}
{(u-w_{1,r_{n+1}})W_n}-\\[6pt]
\hskip10mm- {\tsum_{R_n,S_{n+1}}}\Bigl(\lambda_2(u)+
{\tsum_{r_{n+1}}}\dfrac1{w_{2,s_{n+1}}-w_{1,r_{n+1}}}-
{\tsum_{s_{n+2}\neq s_{n+1}}}\dfrac2{w_{2,s_{n+1}}-w_{2,s_{n+2}}}\times\\[6pt]
\hskip40mm\times
\dfrac{\ket{\MB{w}_1-w_{1,R_n},\MB{w}_2-w_{2,S_n},w_{1,R_n}}}
{(u-w_{2,s_{n+1}})W_n}-\\[6pt]
\hskip10mm- {\tsum_{R_n,S_{n+1}}}~{\tsum_{k=1}^n}
\Bigl(\dfrac2{w_{2,s_{n+1}}-w_{2,s_k}}-
\dfrac2{w_{2,s_{n+1}}-w_{1,r_k}}\Bigr)
\dfrac{\ket{\MB{w}_1-w_{1,R_n},\MB{w}_2-w_{2,S_n},w_{1,R_n}}}
{(u-w_{2,s_{n+1}})W_n}-\\[6pt]
\hskip10mm- {\tsum_{R_n,S_n}}~{\tsum_{k=1}^n} \Bigl(\lambda_3(u)-
{\tsum_{r_{n+1}\neq r_k}}\dfrac1{w_{1,r_k}-w_{1,r_{n+1}}}-
{\tsum_{s_{n+1}\neq
s_k}}\dfrac1{w_{1,r_k}-w_{2,s_{n+1}}}\Bigr)\times\\[6pt]
\hskip40mm\times
\dfrac{\ket{\MB{w}_1-w_{1,R_n},\MB{w}_2-w_{2,S_n},w_{1,R_n}}}
{(u-w_{1,r_k})W_n}+\\[6pt]
\hskip10mm+ {\tsum_{R_n,S_n}}~{\tsum_{{k,\ell=1}\atop{k\neq\ell}}^n}
\Bigl(\dfrac1{w_{1,r_k}-w_{1,r_\ell}}-
\dfrac1{w_{1,r_k}-w_{2,s_\ell}}\Bigr)
\dfrac{\ket{\MB{w}_1-w_{1,R_n},\MB{w}_2-w_{2,S_n},w_{1,R_n}}}
{(u-w_{1,r_k})W_n}+\\[6pt]
\hskip10mm+ {\tsum_{R_{n+1},S_n}}~{\tsum_{k=1}^n}
\dfrac{(w_{2,s_k}-w_{1,r_k})\ket{\MB{w}_1-w_{1,R_n},
\MB{w}_2-w_{2,S_n},w_{1,R_n}}}
{(u-w_{1,r_{n+1}})(u-w_{1,r_k})(w_{2,s_k}-w_{1,r_{n+1}})W_n}=\\[9pt]
\hskip5mm= \tau(u){\tsum_{R_n,S_n}}
\dfrac{\ket{\MB{w}_1-w_{1,R_n},\MB{w}_2-w_{2,S_n},w_{1,R_n}}}{W_n}-\\[6pt]
\hskip10mm- {\tsum_{R_{n+1},S_n}}\Bigl(\lambda_1(u)-
{\tsum_{r_{n+2}\neq r_{n+1}}}\dfrac2{w_{1,r_{n+1}}-w_{1,r_{n+2}}}+
{\tsum_{s_{n+1}}}\dfrac1{w_{1,r_{n+1}}-w_{2,s_{n+1}}}\Bigr)\times\\[6pt]
\hskip40mm\times
\dfrac{\ket{\MB{w}_1-w_{1,R_n},\MB{w}_2-w_{2,S_n},w_{1,R_n}}}
{(u-w_{1,r_{n+1}})W_n}-\\[6pt]
\hskip10mm-
{\tsum_{R_n,S_{n+1}}}\Bigl(\lambda_2(u)+
{\tsum_{r_{n+1}}}\dfrac1{w_{2,s_{n+1}}-w_{1,r_{n+1}}}-
{\tsum_{s_{n+2}\neq s_{n+1}}}\dfrac2{w_{2,s_{n+1}}-w_{2,s_{n+2}}}\times\\[6pt]
\hskip40mm\times
\dfrac{\ket{\MB{w}_1-w_{1,R_n},\MB{w}_2-w_{2,S_n},w_{1,R_n}}}
{(u-w_{2,s_{n+1}})W_n}-\\[6pt]
\hskip10mm- {\tsum_{R_n,S_n}}~{\tsum_{k=1}^n} \Bigl(\lambda_3(u)-
{\tsum_{r_{n+1}\neq r_k}}\dfrac1{w_{1,r_k}-w_{1,r_{n+1}}}-
{\tsum_{s_{n+1}\neq
s_k}}\dfrac1{w_{1,r_k}-w_{2,s_{n+1}}}\Bigr)\times\\[6pt]
\hskip40mm\times
\dfrac{\ket{\MB{w}_1-w_{1,R_n},\MB{w}_2-w_{2,S_n},w_{1,R_n}}}
{(u-w_{1,r_k})W_n}+\\[6pt]
\hskip10mm+ {\tsum_{R_{n+1},S_n}}~{\tsum_{k=1}^n}
\dfrac{(w_{2,s_k}-w_{1,r_k})\ket{\MB{w}_1-w_{1,R_n},
\MB{w}_2-w_{2,S_n},w_{1,R_n}}}
{(u-w_{1,r_k})(w_{1,r_k}-w_{1,r_{n+1}})(w_{2,s_k}-w_{1,r_{n+1}})W_n}-\\[6pt]
\hskip10mm- 2{\tsum_{R_n,S_{n+1}}}~{\tsum_{k=1}^n}
\dfrac{(w_{2,s_k}-w_{2,r_k})\ket{\MB{w}_1-w_{1,R_n},\MB{w}_2-w_{2,S_n},w_{1,R_n}}}
{(u-w_{2,s_{n+1}})(w_{2,s_{n+1}}-w_{2,s_k})(w_{2,s_{n+1}}-w_{1,r_k})W_n}-\\[6pt]
\hskip10mm- {\tsum_{R_n,S_n}}~{\tsum_{{k,\ell=1}\atop{k\neq\ell}}^n}
\dfrac{(w_{2,s_\ell}-w_{1,r_\ell})\ket{\MB{w}_1-w_{1,R_n},\MB{w}_2-w_{2,S_n},w_{1,R_n}}}
{(u-w_{1,r_k})(w_{1,r_k}-w_{2,s_\ell})(w_{1,r_k}-w_{1,r_\ell})W_n}
\end{array}
$$

Hence, we get
$$
\begin{array}{l}
\Bigl(I_0^0(u)-\tau(u;\MB{w}_1,\MB{w}_2)\Bigr)
\MB{P}^n\ket{\MB{w}_1,\MB{w}_2,0}=\\[4pt]
\hskip5mm= {\tsum_{R_n,S_n}}~{\tsum_{k=1}^n}\Bigl(\lambda_1(u)-
{\tsum_{r_{n+1}\neq r_k}}\dfrac2{w_{1,r_k}-w_{1,r_{n+1}}}+
{\tsum_{s_{n+1}}}\dfrac1{w_{1,r_k}-w_{2,s_{n+1}}}\Bigr)\times\\[9pt]
\hskip40mm\times
\dfrac{\ket{\MB{w}_1-w_{1,R_n},\MB{w}_2-w_{2,S_n},w_{1,R_n}}}
{(u-w_{1,r_k})W_n}+\\[9pt]
\hskip10mm+ {\tsum_{R_n,S_n}}~{\tsum_{k=1}^n}\Bigl(\lambda_2(u)+
{\tsum_{r_{n+1}}}\dfrac1{w_{2,s_k}-w_{1,r_{n+1}}}-
{\tsum_{s_{n+1}\neq s_k}}\dfrac2{w_{2,s_k}-w_{2,s_{n+1}}}\times\\[9pt]
\hskip40mm\times
\dfrac{\ket{\MB{w}_1-w_{1,R_n},\MB{w}_2-w_{2,S_n},w_{1,R_n}}}
{(u-w_{2,s_k})W_n}-\\[9pt]
\hskip10mm- {\tsum_{R_n,S_n}}~{\tsum_{k=1}^n} \Bigl(\lambda_3(u)-
{\tsum_{r_{n+1}\neq r_k}}\dfrac1{w_{1,r_k}-w_{1,r_{n+1}}}-
{\tsum_{s_{n+1}\neq
s_k}}\dfrac1{w_{1,r_k}-w_{2,s_{n+1}}}\Bigr)\times\\[9pt]
\hskip40mm\times
\dfrac{\ket{\MB{w}_1-w_{1,R_n},\MB{w}_2-w_{2,S_n},w_{1,R_n}}}
{(u-w_{1,r_k})W_n}+\\[9pt]
\hskip10mm+ {\tsum_{R_{n+1},S_n}}~{\tsum_{k=1}^n}
\dfrac{(w_{2,s_k}-w_{1,r_k})\ket{\MB{w}_1-w_{1,R_n},
\MB{w}_2-w_{2,S_n},w_{1,R_n}}}
{(u-w_{1,r_k})(w_{1,r_k}-w_{1,r_{n+1}})(w_{2,s_k}-w_{1,r_{n+1}})W_n}-\\[9pt]
\hskip10mm- 2{\tsum_{R_n,S_{n+1}}}~{\tsum_{k=1}^n}
\dfrac{(w_{2,s_k}-w_{2,r_k})\ket{\MB{w}_1-w_{1,R_n},\MB{w}_2-w_{2,S_n},w_{1,R_n}}}
{(u-w_{2,s_{n+1}})(w_{2,s_{n+1}}-w_{2,s_k})(w_{2,s_{n+1}}-w_{1,r_k})W_n}-\\[9pt]
\hskip10mm- {\tsum_{R_n,S_n}}~{\tsum_{{k,\ell=1}\atop{k\neq\ell}}^n}
\dfrac{(w_{2,s_\ell}-w_{1,r_\ell})\ket{\MB{w}_1-w_{1,R_n},\MB{w}_2-w_{2,S_n},w_{1,R_n}}}
{(u-w_{1,r_k})(w_{1,r_k}-w_{2,s_\ell})(w_{1,r_k}-w_{1,r_\ell})W_n}
\end{array}
$$
In the same way after reordering we can write
$$
\begin{array}{l}
\tfrac1n\Bigl(I_0^0(u)-\tau(u;\MB{w}_1,\MB{w}_2)\Bigr)
\MB{P}^n\ket{\MB{w}_1,\MB{w}_2,0}=\\[4pt]
\hskip5mm= {\tsum_{R_n,S_n}}\Bigl(\lambda_1(u)- {\tsum_{r_{n+1}\neq
r_n}}\dfrac2{w_{1,r_n}-w_{1,r_{n+1}}}+
{\tsum_{s_{n+1}}}\dfrac1{w_{1,r_n}-w_{2,s_{n+1}}}\Bigr)\times\\[9pt]
\hskip40mm\times
\dfrac{\ket{\MB{w}_1-w_{1,R_n},\MB{w}_2-w_{2,S_n},w_{1,R_n}}}
{(u-w_{1,r_n})W_n}+\\[9pt]
\hskip10mm+ {\tsum_{R_n,S_n}}\Bigl(\lambda_2(u)+
{\tsum_{r_{n+1}}}\dfrac1{w_{2,s_n}-w_{1,r_{n+1}}}-
{\tsum_{s_{n+1}\neq s_n}}\dfrac2{w_{2,s_n}-w_{2,s_{n+1}}}\times\\[9pt]
\hskip40mm\times
\dfrac{\ket{\MB{w}_1-w_{1,R_n},\MB{w}_2-w_{2,S_n},w_{1,R_n}}}
{(u-w_{2,s_n})W_n}-\\[9pt]
\hskip10mm- {\tsum_{R_n,S_n}}\Bigl(\lambda_3(u)-
{\tsum_{r_{n+1}\neq r_n}}\dfrac1{w_{1,r_n}-w_{1,r_{n+1}}}-
{\tsum_{s_{n+1}\neq s_n}}
\dfrac1{w_{1,r_n}-w_{2,s_{n+1}}}\Bigr)\times\\[9pt]
\hskip40mm\times
\dfrac{\ket{\MB{w}_1-w_{1,R_n},\MB{w}_2-w_{2,S_n},w_{1,R_n}}}
{(u-w_{1,r_n})W_n}+\\[9pt]
\hskip10mm+ {\tsum_{R_{n+1},S_n}}
\dfrac{\ket{\MB{w}_1-w_{1,R_n},\MB{w}_2-w_{2,S_n},w_{1,R_n}}}
{(u-w_{1,r_n})(w_{1,r_n}-w_{1,r_{n+1}})(w_{2,s_n}-w_{1,r_{n+1}})W_{n-1}}-\\[9pt]
\hskip10mm- 2{\tsum_{R_n,S_{n+1}}}
\dfrac{\ket{\MB{w}_1-w_{1,R_n},\MB{w}_2-w_{2,S_n},w_{1,R_n}}}
{(u-w_{2,s_{n+1}})(w_{2,s_{n+1}}-w_{2,s_n})(w_{2,s_{n+1}}-w_{1,r_n})W_{n-1}}-\\[9pt]
\hskip10mm- {\tsum_{R_n,S_n}}~{\tsum_{\ell=1}^{n-1}}
\dfrac{(w_{2,s_\ell}-w_{1,r_\ell})\ket{\MB{w}_1-w_{1,R_n},\MB{w}_2-w_{2,S_n},w_{1,R_n}}}
{(u-w_{1,r_n})(w_{1,r_n}-w_{2,s_\ell})(w_{1,r_n}-w_{1,r_\ell})W_n}
\end{array}
$$
If we now use the fact $\lambda_3(u)=\lambda_1(u)+\lambda_2(u)$, we
get after an arrangement
$$
\begin{array}{l}
\Bigl(\tfrac1n\bigl(I_0^0(u)-\tau(u;\MB{w}_1,\MB{w}_2)\bigr)\MB{P}^n
+I_0^+(u)\MB{P}^{n-1}\Bigr)\ket{\MB{w}_1,\MB{w}_2,0}=\\[4pt]
\hskip5mm= -{\tsum_{R_n,S_n}}\Bigl(\lambda_2(w_{2,s_n})+
{\tsum_{r_{n+1}\neq r_n}}\dfrac2{w_{1,r_n}-w_{1,r_{n+1}}}-
{\tsum_{s_{n+1}}}\dfrac1{w_{1,r_n}-w_{2,s_{n+1}}}\Bigr)\times\\[9pt]
\hskip40mm\times
\dfrac{\ket{\MB{w}_1-w_{1,R_n},\MB{w}_2-w_{2,S_n},w_{1,R_n}}}
{(u-w_{1,r_n})W_n}+\\[9pt]
\hskip10mm+ {\tsum_{R_n,S_n}}\Bigl(\lambda_2(w_{2,s_n})+
{\tsum_{r_{n+1}}}\dfrac1{w_{2,s_n}-w_{1,r_{n+1}}}-
{\tsum_{s_{n+1}\neq s_n}}\dfrac2{w_{2,s_n}-w_{2,s_{n+1}}}\Bigr)\times\\[9pt]
\hskip40mm\times
\dfrac{\ket{\MB{w}_1-w_{1,R_n},\MB{w}_2-w_{2,S_n},w_{1,R_n}}}
{(u-w_{2,s_n})W_n}+\\[9pt]
\hskip10mm+ {\tsum_{R_n,S_n}}\Bigl({\tsum_{r_{n+1}\neq r_n}}
\dfrac1{w_{1,r_n}-w_{1,r_{n+1}}}+ {\tsum_{s_{n+1}\neq s_n}}
\dfrac1{w_{1,r_n}-w_{2,s_{n+1}}}\Bigr)\times\\[9pt]
\hskip40mm\times
\dfrac{\ket{\MB{w}_1-w_{1,R_n},\MB{w}_2-w_{2,S_n},w_{1,R_n}}}
{(u-w_{1,r_n})W_n}+\\[9pt]
\hskip10mm+ {\tsum_{R_{n+1},S_n}}
\dfrac{\ket{\MB{w}_1-w_{1,R_n},\MB{w}_2-w_{2,S_n},w_{1,R_n}}}
{(u-w_{1,r_n})(w_{1,r_n}-w_{1,r_{n+1}})(w_{2,s_n}-w_{1,r_{n+1}})W_{n-1}}+\\[9pt]
\hskip10mm+ 2{\tsum_{R_n,S_{n+1}}}
\dfrac{\ket{\MB{w}_1-w_{1,R_n},\MB{w}_2-w_{2,S_n},w_{1,R_n}}}
{(u-w_{1,r_n})(w_{1,r_n}-w_{2,s_{n+1}})(w_{2,s_{n+1}}-w_{2,s_n})W_{n-1}}-\\[9pt]
\hskip10mm- {\tsum_{R_n,S_n}}~{\tsum_{\ell=1}^{n-1}}
\dfrac{(w_{2,s_\ell}-w_{1,r_\ell})\ket{\MB{w}_1-w_{1,R_n},\MB{w}_2-w_{2,S_n},w_{1,R_n}}}
{(u-w_{1,r_n})(w_{1,r_n}-w_{2,s_\ell})(w_{1,r_n}-w_{1,r_\ell})W_n}
\end{array}
$$
We now introduce $\lambda_2(w_{2,s_n})$ from (\ref{B2}) in the above
expression and obtain
$$
\begin{array}{l}
\Bigl(\tfrac1n\bigl(I_0^0(u)-\tau(u;\MB{w}_1,\MB{w}_2)\bigr)\MB{P}^n
+I_0^+(u)\MB{P}^{n-1}\Bigr)\ket{\MB{w}_1,\MB{w}_2,0}=\\[4pt]
\hskip5mm= -{\tsum_{R_n,S_n}}\biggl[{\tsum_{r_{n+1}\neq r_n}}
\Bigl(\dfrac1{w_{1,r_n}-w_{1,r_{n+1}}}-
\dfrac1{w_{2,s_n}-w_{1,r_{n+1}}}\Bigr)+\\[9pt]
\hskip15mm+ {\tsum_{s_{n+1}\neq s_n}}
\Bigl(\dfrac2{w_{2,s_n}-w_{2,s_{n+1}}}-
\dfrac2{w_{1,r_n}-w_{2,s_{n+1}}}\Bigr)\biggr]
\dfrac{\ket{\MB{w}_1-w_{1,R_n},\MB{w}_2-w_{2,S_n},w_{1,R_n}}}
{(u-w_{1,r_n})W_n}+\\[9pt]
\hskip10mm+ {\tsum_{R_{n+1},S_n}}
\dfrac{\ket{\MB{w}_1-w_{1,R_n},\MB{w}_2-w_{2,S_n},w_{1,R_n}}}
{(u-w_{1,r_n})(w_{1,r_n}-w_{1,r_{n+1}})(w_{2,s_n}-w_{1,r_{n+1}})W_{n-1}}+\\[9pt]
\hskip10mm+ 2{\tsum_{R_n,S_{n+1}}}
\dfrac{\ket{\MB{w}_1-w_{1,R_n},\MB{w}_2-w_{2,S_n},w_{1,R_n}}}
{(u-w_{1,r_n})(w_{1,r_n}-w_{2,s_{n+1}})(w_{2,s_{n+1}}-w_{2,s_n})W_{n-1}}-\\[9pt]
\hskip10mm- {\tsum_{R_n,S_n}}~{\tsum_{k=1}^{n-1}}
\dfrac{(w_{2,s_k}-w_{1,r_k})\ket{\MB{w}_1-w_{1,R_n},\MB{w}_2-w_{2,S_n},w_{1,R_n}}}
{(u-w_{1,r_n})(w_{1,r_n}-w_{2,s_k})(w_{1,r_n}-w_{1,r_k})W_n}=\\[12pt]
\hskip5mm= -{\tsum_{R_n,S_n}}~{\tsum_{r_{n+1}\neq
r_n}}\dfrac{\ket{\MB{w}_1-w_{1,R_n},\MB{w}_2-w_{2,S_n},w_{1,R_n}}}
{(u-w_{1,r_n})(w_{1,r_n}-w_{1,r_{n+1}})(w_{2,s_n}-w_{1,r_{n+1}})W_{n-1}}+\\[9pt]
\hskip10mm+ {\tsum_{R_{n+1},S_n}}
\dfrac{\ket{\MB{w}_1-w_{1,R_n},\MB{w}_2-w_{2,S_n},w_{1,R_n}}}
{(u-w_{1,r_n})(w_{1,r_n}-w_{1,r_{n+1}})(w_{2,s_n}-w_{1,r_{n+1}})W_{n-1}}+\\[9pt]
\hskip10mm+ 2{\tsum_{R_n,S_n}}~{\tsum_{s_{n+1}\neq s_n}}
\dfrac{\ket{\MB{w}_1-w_{1,R_n},\MB{w}_2-w_{2,S_n},w_{1,R_n}}}
{(u-w_{1,r_n})(w_{1,r_n}-w_{2,s_{n+1}})(w_{2,s_n}-w_{2,s_{n+1}})W_{n-1}}+\\[9pt]
\hskip10mm+ 2{\tsum_{R_n,S_{n+1}}}
\dfrac{\ket{\MB{w}_1-w_{1,R_n},\MB{w}_2-w_{2,S_n},w_{1,R_n}}}
{(u-w_{1,r_n})(w_{1,r_n}-w_{2,s_{n+1}})(w_{2,s_{n+1}}-w_{2,s_n})W_{n-1}}-\\[9pt]
\hskip10mm- {\tsum_{R_n,S_n}}~{\tsum_{k=1}^{n-1}}
\dfrac{(w_{2,s_k}-w_{1,r_k})\ket{\MB{w}_1-w_{1,R_n},\MB{w}_2-w_{2,S_n},w_{1,R_n}}}
{(u-w_{1,r_n})(w_{1,r_n}-w_{2,s_k})(w_{1,r_n}-w_{1,r_k})W_n}
\end{array}
$$
It can be rewritten in the form
$$
\begin{array}{l}
\Bigl(\tfrac1n\bigl(I_0^0(u)-\tau(u;\MB{w}_1,\MB{w}_2)\bigr)\MB{P}^n
+I_0^+(u)\MB{P}^{n-1}\Bigr)\ket{\MB{w}_1,\MB{w}_2,0}=\\[4pt]
= -{\tsum_{R_n,S_n}}~{\tsum_{k=1}^{n-1}}\Bigl(
\dfrac1{(w_{1,r_n}-w_{1,r_k})(w_{2,s_n}-w_{1,r_k})(w_{2,s_k}-w_{1,r_k})}+\\[9pt]
\hskip25mm+
\dfrac2{(w_{2,s_n}-w_{2,s_k})(w_{2,s_k}-w_{1,r_n})(w_{2,s_k}-w_{1,r_k})}-\\[9pt]
\hskip25mm-
\dfrac1{(w_{1,r_n}-w_{1,r_k})(w_{2,s_k}-w_{1,r_n})(w_{2,s_n}-w_{1,r_n})}\Bigr)\times\\[12pt]
\hskip35mm\times
\dfrac{(w_{2,s_n}-w_{1,r_n})(w_{2,s_k}-w_{1,r_k})}{W_n}
\ket{\MB{w}_1-w_{1,R_n},\MB{w}_2-w_{2,S_n},w_{1,R_n}}.
\end{array}
$$
Now we see that the elements in the bracket are antisymmetric under
the change of $s_k$ by $s_n$ and the second term is symmetric. So we
have
$$
\Bigl(\tfrac1n\bigl(I_0^0(u)-\tau(u;\MB{w}_1,\MB{w}_2)\bigr)\MB{P}^n
+I_0^+(u)\MB{P}^{n-1}\Bigr)\ket{\MB{w}_1,\MB{w}_2,0}=0\,.
$$

\section{Concluding remarks and open problems}
In the present paper we have proposed the new formula for the
eigenvectors of the Gaudin model obtained by using the Bethe ansatz
method in the $\sl(3)$ case.

In the the $\sl(3)$ case we can use the formula (\ref{F1}) for these
eigenvectors from the paper \cite{Feigin}. The first interesting
problem is to find an explicit connection. We were able to reduce
one to the other only in some simple examples such as:
$$
\left|\,w_{1,1},w_{2,1}\right)= \Bigl(F_1(w_{1,1})F_2(w_{2,1})+
\frac{F_3(w_{1,1})}{w_{2,1}-w_{1,1}}\Bigr)\ket{0}
$$
and
$$
\begin{array}{l}
\left|\,{w}_{1,1},{w}_{1,2},{w}_{2,1},{w}_{2,2}\right)=
\Bigl( F_1(w_{1,1})F_1(w_{1,2})F_2(w_{2,1})F_2(w_{2,2})+\\[6pt]
\hskip20mm+
\dfrac{F_1(w_{1,1})F_2(w_{2,1})F_3(w_{1,2})}{w_{2,2}-w_{1,2}}+
\dfrac{F_1(w_{1,1})F_2(w_{2,2})F_3(w_{1,2})}{w_{2,1}-w_{1,2}}+\\[9pt]
\hskip30mm+
\dfrac{F_1(w_{1,2})F_2(w_{2,1})F_3(w_{1,1})}{w_{2,2}-w_{1,1}}+
\dfrac{F_1(w_{1,2})F_2(w_{2,2})F_3(w_{1,1})}{w_{2,1}-w_{1,1}}+\\[9pt]
\hskip40mm+
\dfrac{F_3(w_{1,1})F_3(w_{1,2})}{(w_{2,2}-w_{1,2})(w_{2,1}-w_{1,1})}+
\dfrac{F_3(w_{1,1})F_3(w_{1,2})}{(w_{2,1}-w_{1,2})(w_{2,2}-w_{1,1})}
\Bigr) \ket{0}
\end{array}
$$
but we believe that it is possible generally. We studied the case of
the algebra $\sl(3)$ explicitly. We believe that similar formulas
are possible for the general semisimple Lie algebra. Some
calculation for the $B_2$ algebra is in progress. So the second open
problem is to generalize our method to other Lie algebras.

All proofs in the presented paper are direct calculations. So the
last problem is  to find some indirect proof which can be useful in
the general case.

\section*{Acknowledgement}

This work was partially supported by research plan MSM6840770039.

\end{document}